\documentclass[superscriptaddress,twocolumn,pre]{revtex4-1}

\usepackage{graphicx}      
\usepackage{hyperref}      
\usepackage{amsmath,amssymb} 
\usepackage{mathrsfs}
\usepackage{multirow}      
\usepackage{xcolor}        
\usepackage{pgfplots}      
\usepackage{lipsum}        
\usepackage{comment}

\graphicspath{{figs/}}

\DeclareMathAlphabet{\mathpzc}{OT1}{pzc}{m}{it}

\begin{document}
\author{Fatemeh Aghaei}
	\affiliation{Max Planck Institute for the Physics of Complex Systems, 01187 Dresden, Germany}
 \author{Abbas Ali Saberi}
\email{corresponding author: asaberi@constructor.university}
\affiliation{School of Science, Constructor University, Campus Ring 1, 28759 Bremen,
Germany}
\affiliation{Max Planck Institute for the Physics of Complex Systems, 01187 Dresden, Germany}
 \author{Holger Kantz}
		\affiliation{Max Planck Institute for the Physics of Complex Systems, 01187 Dresden, Germany}
\author{J{\"u}rgen Kurths}
 \affiliation{Potsdam Institute for Climate Impact Research, Potsdam, Germany} 
 \affiliation{Department of Physics, Humboldt University, Berlin, Germany}
	
\title{Superstable Geometry in Triadic Percolation}

\begin{abstract}
Triadic percolation turns bond percolation into a dynamical problem governed by an effective one-dimensional unimodal map. We show that the geometry of superstable cycles provides a direct, map-agnostic probe of local nonlinearity: specifically, the distance from the map’s maximum to a distinguished next-to-maximum point on the attracting $2^n$-cycle (which coincides with a preimage of the maximum at $2^n$-superstability) scales as $|\Delta p|^{\gamma}$ with $\gamma = 1/z$, where $z$ is the nonflat order of the maximum. This prediction is verified across canonical unimodal families and heterogeneous triadic ensembles, with Lyapunov spectra corroborating the one-dimensional reduction. A derivative condition on the activation kernel fixes the local nonlinearity order $z$ (and thus, under standard unimodal-map hypotheses, the associated $z$-logistic universality class) and gives conditions under which $z>2$ can be realized. The diagnostic operates directly on orbit data under standard regularity assumptions, providing a practical tool to classify universality in higher-order networks.
\end{abstract}

\maketitle

\section{Introduction} 

Percolation theory is one of the simplest paradigms in statistical physics for connectivity–driven phase transitions: as an occupation probability crosses a critical threshold, a giant connected component emerges and controls large–scale behavior \cite{StaufferAharony1994,Saberi2015PhysRep}. Despite its minimal rules, percolation has proved remarkably versatile, underpinning models of transport in disordered media, geomorphology, infrastructure robustness, and complex networks more broadly \cite{Saberi2015PhysRep,NewmanStrogatzWatts2001}. Triadic percolation builds directly on this foundation and provides a minimal setting for higher–order network dynamics in which nodes regulate the activation of links and, in turn, node activity emerges from connectivity through the giant component of active links. This regulation–percolation feedback turns percolation into a genuinely dynamical process that can display fixed points, period–doubling cascades, and chaos across heterogeneous topologies and regulator statistics \cite{Sun2023NatComm,Millan2024PNASNexus,Sun2024PRE, sun2026triadic}. Such phenomena arise broadly in systems where interactions go beyond pairs—ranging from biological and social systems to technological infrastructures—motivating models that explicitly encode higher–order structure \cite{Battiston2020PhysRep,Battiston2021NatPhys,boccaletti2023structure,bairey2016high,grilli2017higher,niedostatek2025mining}. A key challenge, however, is universality identification when the effective one-dimensional map remains implicit or experimentally inaccessible: standard approaches based on Feigenbaum ratios or delicate parameter-scaling fits \cite{Feigenbaum1978,Feigenbaum1979} are often impractical or fragile in data-driven settings. This motivates a direct, geometric, map-agnostic route to universality that operates on orbit diagrams themselves---i.e., directly on the measured state-parameter geometry (windows, superstable points, branch structure)---without requiring an explicit functional form.

Geometry is central to triadic percolation at two levels: at the network level, higher-order structure (nodes regulating links) shapes the effective interaction topology; at the dynamical level, the state--parameter geometry of iterates (fixed points, periodic windows, bifurcation trees, superstable cycles) organizes routes to complexity. In regulation–percolation maps, the local shape near the maximum (the critical point of the effective one-dimensional map) controls the onset of period doubling and chaos, while global patterns---window arrangement, superstable spacing, and laminar-phase scaling---yield robust geometric signatures insensitive to microscopic details. Classic unimodal studies have long leveraged such structure via bifurcation diagrams, kneading sequences, Lyapunov ridges, and Feigenbaum-type scalings to classify dynamics without explicit functional forms \cite{MilnorThurston1988,ColletEckmann1980,DeMeloVanStrien1993,Feigenbaum1978,Feigenbaum1979}. In higher-order network settings, geometry-driven diagnostics similarly operate directly on data, enabling comparisons across models and experiments without reconstructing governing equations \cite{Battiston2020PhysRep,boccaletti2023structure}. Thus geometry provides a natural, practical lens for triadic percolation, linking local nonlinearity to global organization and supporting model-agnostic inference in heterogeneous, data-rich contexts.

Within this geometric lens, superstable geometry is a sharp probe. A superstable cycle contains the critical point and thus has zero multiplier, making it a structural landmark that anchors the bifurcation tree in state and parameter space. Its positions organize periodic windows and the backbone of period doubling, and they serve as noise-robust markers in orbit diagrams because the local derivative vanishes at the critical point. In higher-order network settings---where the effective one-dimensional map is often implicit---superstable geometry operates directly on trajectories, enables comparison across heterogeneous topologies and regulator statistics, and ties observable state–parameter patterns to the nonlinear shape that governs the onset of complex behavior.
Related directions on percolation, resilience, and dynamics in complex (including higher-order and quantum) networks include
Refs.~\cite{Dong2021PNAS_ModularResilience, Zhang2024SciAdv_HigherOrderBasins, Metz2025PRL_DMFT_SparseDirected, Hu2025SciAdv_NonshortestQuantumPaths}.

Here we show how to extract the governing nonlinearity directly from orbit diagrams of triadic percolation. We introduce a map-agnostic diagnostic based on superstable geometry that works on data without reconstructing the effective map, and we give an analytic derivation that links the local shape at the dynamical maximum to a measurable scaling exponent $\gamma=1/z$. We then validate the diagnostic on standard unimodal families and apply it to triadic percolation with varied regulator and structural ensembles, confirm the effective one-dimensional dynamics via Lyapunov-spectrum checks, and test robustness to finite sampling and noise. Finally, we specify the parameter and data regimes where the diagnostic is reliable and discuss implications for higher-order network dynamics and data-driven analysis. Beyond the probability–generating–function (PGF) specification of the degree distributions used for the microscopic model, we also analyze a phenomenological variant in which the regulator–to–link response is modeled by smooth “effective kernels” (Hill–type functions with possibly noninteger exponents).

\section{Method}

\paragraph*{Network structure.}
We consider a structural network $G=(V,E)$, where $V$ and $E$ are sets of vertices and edges respectively, and a signed regulatory layer $W=W^+\cup W^-$ on the same node set $V$. Each directed regulatory edge $(i\!\to\!\ell)\in W^\pm$ points from a node $i\in V$ to a structural link $\ell\in E$ and encodes positive ($+$) or negative ($-$) regulation of that link. The structural node degrees $k$ are drawn i.i.d.\ from $\pi(k)$ with mean $\langle k\rangle$, independently of regulation. For the regulatory layer, each node has independent out--degrees $\hat{\kappa}^\pm$ drawn from $\hat{P}^\pm(\hat{\kappa}^\pm)$. Opposite--sign regulation from the same node to the same link is forbidden. Networks are sparse and locally tree--like \cite{Sun2023NatComm}.

\paragraph*{Dynamics (triadic percolation).}
Time is discrete. Let an undirected structural link be \emph{active} at time $t$ with probability $p_L^{(t)}$. Given the active–link subgraph, a node is active if it belongs to the giant connected component (GCC) of active links. Regulation updates structural link states as follows. A link is \emph{forbidden} at time $t$ if at least one active negative regulator targets it or if no active positive regulator targets it; forbidden links are set to be inactive. All other structural links are \emph{admissible}: an admissible inactive link becomes active with probability $p$ and remains inactive with probability $q=1-p$. Updates are synchronous. This setup follows Refs.~\cite{Sun2023NatComm,Millan2024PNASNexus,Sun2024PRE}.

\paragraph*{Mean--field equations.}
Under the locally tree--like assumption, the dynamics are captured by the closed recursion for $S^{(t)}$, $R^{(t)}$, and $p_L^{(t)}$:
\begin{equation}
\begin{aligned}
S^{(t)} &= 1 - G_1\!\left(1 - S^{(t)}\,p_L^{(t-1)}\right),\\
R^{(t)} &= 1 - G_0\!\left(1 - S^{(t)}\,p_L^{(t-1)}\right),\\
p_L^{(t)} &= p\,G_0^{-}\!\left(1 - R^{(t)}\right)\Bigl[1 - G_0^{+}\!\left(1 - R^{(t)}\right)\Bigr],
\end{aligned}
\end{equation}
where $S^{(t)}$ is the probability that a node at the end of a randomly chosen structural link belongs to the GCC at time $t$, $R^{(t)}$ is the probability that a node at the end of a randomly chosen regulatory edge belongs to the GCC at time $t$, and $p_L^{(t)}$ is the activation probability of a structural link at time $t$. The generating functions are
\begin{equation}
\begin{aligned}
G_0(x) &= \sum_k \pi(k)\,x^k,\\
G_1(x) &= \sum_k \pi(k)\,\frac{k}{\langle k\rangle}\,x^{k-1},\\
G_0^{\pm}(x) &= \sum_{\hat{\kappa}^\pm}\hat{P}^\pm(\hat{\kappa}^\pm)\,x^{\hat{\kappa}^\pm},
\end{aligned}
\end{equation}
with $G_0$ and $G_1$ as in the standard configuration-model percolation formalism, which are related by $G'_0 = \langle k\rangle G_1$ \cite{NewmanStrogatzWatts2001,Newman2002}. The control parameter is $p\in[0,1]$; the structural and regulatory degree distributions enter only through $G_0$, $G_1$, and $G_0^\pm$.

\paragraph*{Observables and protocol.}
We take $R^{(t)}$ as the order parameter and work with the effective one–dimensional map obtained by composing the regulation step with bond percolation. Define the regulatory kernel
\begin{equation}
\Phi_p(R) \;:=\; p\,G_0^{-}(1-R)\,\bigl[1-G_0^{+}(1-R)\bigr].
\end{equation}
(equivalently, we may write $\Phi_p(R) \equiv p_L(R)$,
and below we freely use $\Phi_p$ and $p_L$ interchangeably when the dependence on $R$ is clear.)
For any link–activation level $y\in[0,1]$, let $S(y)$ denote the nonzero solution corresponding to the giant-component branch of
\begin{equation}
S(y) \;=\; 1 - G_1\!\bigl(1 - S(y)\,y\bigr).
\label{eq: S}
\end{equation}
Then set the percolation closure
\begin{equation}
\Psi(y) \;:=\; 1 - G_0\!\bigl(1 - S(y)\,y\bigr).
\label{eq: psi}
\end{equation}
The effective map is then
\begin{equation}
R^{(t)} \;=\; H_p\!\bigl(R^{(t-1)}\bigr).
\label{eq: 1D}
\end{equation}
Here we have dropped the explicit time indices for $S$ and $p_L$: given $R^{(t-1)}$, the quantities $S^{(t)}$ and $p_L^{(t)}$ are uniquely determined by Eqs.~\eqref{eq: S} and \eqref{eq: psi}, so the dynamics reduce to iterating the one-dimensional map $R^{(t)} = H_p(R^{(t-1)})$.
With
\begin{equation}
H_p(R)\;=\;\Psi\!\big(\Phi_p(R)\big),
\end{equation}
so by the chain rule
\begin{equation}
H_p'(R)\;=\;\Psi'\!\big(\Phi_p(R)\big)\;\Phi_p'(R),
\end{equation}
where the prime denotes differentiation with respect to $R$. With $f(x):=G_0^-(x)$ and $g(x):=G_0^+(x)$, and $x:=1-R$ we set,
\begin{equation}
\Phi_p(R)\;=\;p\,f(x)\,[1-g(x)],
\end{equation}
and
\begin{equation}
\Phi_p'(R)\;=\;p\Big(-f'(x)[1-g(x)]+f(x)g'(x)\Big).
\end{equation}
From Eq.~\eqref{eq: psi} we obtain
\begin{equation}
\Psi'(y)=G_0'\!\big(1-S(y)\,y\big)\,\big(S'(y)\,y+S(y)\big).
\end{equation}

For readability, we defer the details of the effective-kernel (Hill-type) variant to a later subsection.

To make monotonicity explicit, {we rewrite Eq. \ref{eq: S} and differentiate $\mathcal{F}(S,y):=S-1+G_1(1-Sy)=0$:
\begin{equation}
\frac{\partial \mathcal{F}}{\partial S}\,S'(y)+\frac{\partial \mathcal{F}}{\partial y}=0,
\end{equation}
to reach out the following expression:
\begin{equation}
\bigl[1-y\,G_1'(1-S(y)~y)\bigr]\,S'(y)=G_1'(1-S(y)~y)\,S(y).
\end{equation}
Hence
\begin{equation}
S'(y)=\frac{G_1'(1-S(y)~y)\,S(y)}{1-y\,G_1'(1-S(y)~y)}\;\ge\;0,
\end{equation}
provided $1-y\,G_1'(1-S(y)~y)>0$, which holds on the operating range below the structural percolation instability. Therefore, $\Psi'(y) > 0$
since $G_0'(u)>0$ on $(0,1]$ and $S(y)\in[0,1]$.

Therefore the interior extrema of $H_p$ occur at $\Phi_p'(R)=0$, i.e.
\begin{equation}
\frac{f'(x)}{f(x)}=\frac{g'(x)}{1-g(x)}\qquad \text{with } x=1-R,
\end{equation}
equivalently,
\begin{equation}
\frac{d}{dx}\log F(x)=0 \quad \text{for } F(x)=f(x)\,[1-g(x)].
\end{equation}
The \emph{critical point} (maximum) used in the superstable geometry is
\begin{equation}
R_m := \arg\max_{R\in[0,1]} H_p(R).
\end{equation}
In particular, because $\Psi'>0$ in the regime of interest, $H_p'(R_m)=0$ if and only if $\Phi_p'(R_m)=0$. Thus the location of the maximum $R_m$ is determined by the regulator statistics through the condition $\Phi_p'(R)=0$ above.\\
Equation~\eqref{eq: 1D} yields a one-dimensional iterated map for $R$. We compute the Lyapunov exponent of the $R$-map (after discarding transients) as
\begin{equation}
\lambda = \lim_{T\to\infty}\frac{1}{T}\sum_{t=1}^{T}\ln\!\left|H_p'\!\big(R^{(t-1)}\big)\right|.
\end{equation}
At superstable points the derivative vanishes, so formally $\lambda\to -\infty$; numerically they appear as sharp minima in $\lambda(p)$. 
In practice, the superstable parameters $p_n$ are detected as sharp minima of $\lambda(p)$ in a coarse scan. Each minimum is then refined by a local bracket-and-solve procedure enforcing the superstable condition that the attracting $2^n$-cycle passes through $R_m$
(equivalently, that the $2^n$-return at $R_m$ closes).
For robustness, we also evolve the coupled recursion for $(R^{(t)},S^{(t)})$ and estimate the two largest Oseledets exponents via QR decomposition. In our data, the leading exponent from the 2D embedding agrees with $\lambda$, while the subleading exponent is strongly negative, corroborating an effectively one-dimensional central manifold \cite{Kantz1994,Rosenstein1993,Benettin1980}.

\subsection*{Theory: superstable geometry gives $\gamma=1/z$}

Let $H_p(R)$ be the effective one–dimensional map with a single interior maximum at $R_m$. Assume the first nonzero derivative at $R_m$ is the $z$th one (even $z\ge 2$). Let $p_n$ denote the parameters where the attracting cycle has period $2^n$ and passes through $R_m$ (the $2^n$–superstable points).

Define the $2^n$-step return map $F_p(R):=H_p^{(2^n)}(R)$.
We introduce $\mu_n(p):=F_p(R_m)-R_m$, so that $\mu_n(p_n)=0$ at the $2^n$-superstable point.
Let $z$ be the smallest even integer for which $\partial_R^{\,z}F_{p_n}(R_m)\neq 0$ and define
$A_n := -\frac{1}{z!}\,\partial_R^{\,z}F_{p_n}(R_m) > 0$.
Assuming a nondegenerate unfolding, $\mu_n(p)=\mu_n'(p_n)(p-p_n)+O\!\big((p-p_n)^2\big)$.

Near each $p_n$, consider the $2^n$–step return map $H_p^{(2^n)}(R)$. In a neighborhood of $R_m$,

\begin{equation}
F_p(R_m+\delta)=R_m+\mu_n(p)-A_n|\delta|^{z}+ \cdots,
\end{equation}

with $A_n>0$, $\mu_n(p)\simeq \mu_n'(p_n)(p-p_n)$ and $\mu_n'(p_n)\neq 0$ expresses a generic transversality (nondegenerate unfolding) of the $2^n$-return at $R_m$ as $p$ varies, a standard hypothesis in unimodal dynamics \cite{DeMeloVanStrien1993,ColletEckmann1980} and borne out in our numerics. The return map has the same local nonlinearity order $z$ at $R_m$.

On the period-$2^n$ orbit, for $p$ near but not equal to $p_n$, we define $R_n(p)$ as the solution (on the selected continuous branch) of
$H_p^{(2^n)}(R_n(p))=R_m$.
We then define the corresponding point at superstability by $R_n(p_n):=\lim_{p\to p_n}R_n(p)$, and write $R_n(p)=R_m+\delta_n(p)$.

Solving $H_p^{(2^n)}(\mathcal{R}_n(p))=R_m$ gives 
\begin{equation}
|\delta_n(p)|\;\propto\;|p-p_n|^{1/z}.
\end{equation}
Defining the measured slope by $|\delta_n(p)|\propto|p-p_n|^{\gamma}$ therefore yields
\begin{equation}
\gamma=\frac{1}{z}.
\end{equation}

\subsection*{Superstable geometry: extraction protocol and scaling}

\paragraph*{Branch tracking and alternation.}
For each $n$, and for $p$ in a small interval on one side of $p_n$, we follow the solution $R_n(p)$ by parameter continuation:
starting from a seed at $p=p_n\pm\varepsilon$, we step $p$ on a fine grid and at each step we choose the solution closest to the previous one
(nearest-neighbor continuation). We also keep a fixed sign convention for $\delta_n(p)=R_n(p)-R_m$ (i.e., we do not jump between branches).
This identifies one continuous branch until it ends at the window boundary.
\\
\noindent\rule{\columnwidth}{0.4pt}
\textbf{Algorithm 1: Superstable-geometry diagnostic (implementation summary).}\\
\emph{Input:} parameter grid $\{p\}$; effective map $H_p$; critical point $R_m$ (map maximum).\\
\emph{Output:} scaling exponent $\gamma$ (and fitted slopes $a_i$).
\begin{enumerate}
\item Scan $p$ and compute $\lambda(p)$ from the $R$-map after discarding transients; detect candidate $p_n$ as sharp minima of $\lambda(p)$.
\item For each candidate, refine $p_n$ locally by bracketing + root-finding, such as Newton-Raphson method, of the superstable condition (the attracting $2^n$-cycle passes through $R_m$).
\item For each $n$, choose a small $\varepsilon$ and initialize on one side: set $p=p_n+\varepsilon$ (or $p_n-\varepsilon$) and solve
      $H_p^{(2^n)}(R)=R_m$ for $R$ near $R_m$ to obtain a seed $R_n(p)$.
\item Continue the branch by stepping $p$ monotonically and at each step choose the solution closest to the previous $R_n(p)$
      (nearest-neighbor continuation), keeping a fixed sign convention for $\delta_n(p)=R_n(p)-R_m$.
\item For each $i<n$, compute $\Delta p_{ni}=|p_n-p_i|$ and $d_{ni}=|R_n(p_i)-R_m|$;
      fit $\log d_{ni}$ vs.\ $\log \Delta p_{ni}$ over the chosen fit range to extract $a_i$ and the asymptotic $\gamma$.
\end{enumerate}

\noindent\rule{\columnwidth}{0.4pt}
In practice, from the orbit diagram $R$ vs.\ $p$, locate $R_m$ and the superstable parameters $p_n$. Starting from a reference superstable point $(p_i,R_m)$, we follow the preimage of $R_m$ through successive period doublings: for each $n>i$ we identify on the period–$2^n$ cycle the point $\mathcal{R}_n(p)$ that smoothly continues the preimage of $R_m$ at $p=p_n$. To maintain continuity across windows, we track this point along an alternating up/down subbranch on either side of $R_m$.

Fix a reference $p_i$ on this branch and define
\begin{equation}
d_{ni}=|\mathcal{R}_n(p_n)-\mathcal{R}_n(p_i)|,\qquad
\Delta p_{ni}=|p_n-p_i|\quad(n>i).
\end{equation}
Using the local return–map expansion above,
\begin{equation}
d_{ni}\;\propto\;(\Delta p_{ni})^{\gamma},
\end{equation}
and we estimate $\gamma$ by linear fits of $\log d_{ni}$ vs.\ $\log \Delta p_{ni}$ on contiguous windows where the branch is well resolved.

\section{Results}

\subsection*{Benchmarks on canonical unimodal families}
Applying the extraction to the quadratic unimodal map $f_{\mathrm{quad}}(x)=r-x^2$ yields $\gamma=0.5$ as expected for $z=2$, while a quartic unimodal family ($z=4$) yields $\gamma=0.25$ (Fig.~\ref{fig: LogisticMap}), in agreement with the unimodal theory \cite{DeMeloVanStrien1993,ColletEckmann1980}.

\begin{figure}[h]
\centering
\includegraphics[width=0.48\textwidth]{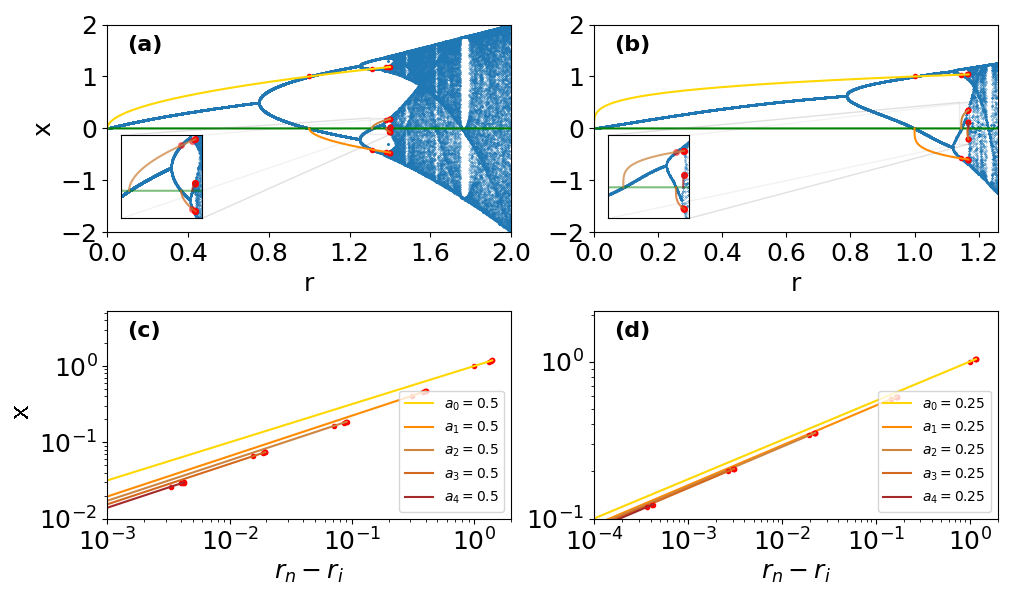}
\caption{Benchmarks on canonical unimodal maps. (a,b) Alternating-branch tracking for $z=2$ and $z=4$. (c,d) Log--log fits return $\gamma=0.5$ (quadratic) and $\gamma=0.25$ (quartic), verifying $\gamma=1/z$. $a_i$ is the slope from a fit of the data points $(\log \Delta p_{ni},\log d_{ni})$ over the chosen fit range (equivalently, the corresponding $i$-range) for the tracked branch/sequence.}
\label{fig: LogisticMap}
\end{figure}
\subsection*{Period doubling, chaos, and effective dimensionality}
Orbit diagrams of $R$ versus $p$ exhibit the fixed point $\to$ period-doubling $\to$ chaos route for both Poisson structural networks with Poisson regulators and scale-free structures with Poisson regulators (Fig.~\ref{fig: TriadicPercolation}a,b). Lyapunov analysis confirms the effective one-dimensional dynamics: the leading exponent $\lambda$ computed from the reduced $R$-map is positive in the chaotic windows, while a two-dimensional embedding with QR decomposition yields a largest exponent $\lambda_{\max}\approx\lambda$ and a strongly negative subleading exponent $\lambda_2\ll 0$ (Fig.~\ref{fig: TriadicPercolation}c,d), consistent with a one-dimensional central manifold \cite{Kantz1994,Benettin1980}. Here, the two-dimensional embedding refers to evolving the coupled recursion for $(R^{(t)},S^{(t)})$ given in the Method (Mean–field equations), and computing the Lyapunov spectrum of that 2D map.\\

\begin{figure}[t]
\centering
\includegraphics[width=0.48\textwidth]{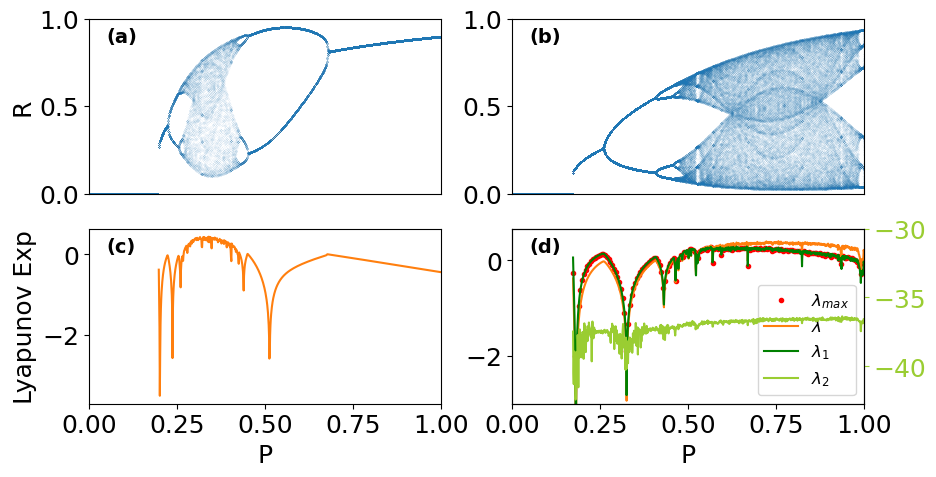}
\caption{Period doubling and chaos in triadic percolation. (a) Poisson structure ($\langle k\rangle=30$) with Poisson regulators ($\langle \hat{\kappa}^+\rangle=1.8$, $\langle \hat{\kappa}^-\rangle=2.5$). (b) Scale-free structure (degree exponent $\gamma_{\mathrm{deg}}=2.5$, $k_{\min}=4$, $k_{\max}=100$) with Poisson regulators ($\langle \hat{\kappa}^+\rangle=10$, $\langle \hat{\kappa}^-\rangle=2.8$). (c,d) Lyapunov spectra: the leading exponent $\lambda>0$ in chaotic windows; a two-dimensional embedding yields $\lambda_1 = \lambda_{\max}\approx\lambda$ and a strongly negative subleading exponent $\lambda_2\ll 0$, indicating effectively one-dimensional dynamics. Here, $\lambda$ is the one-dimensional Lyapunov exponent, $\lambda_{\max}$ is the largest exponent estimated using Kantz's method \cite{Kantz1994}, and $\lambda_1$ is the largest exponent obtained from QR decomposition.}
\label{fig: TriadicPercolation}
\end{figure}

For triadic percolation with Poisson regulators, the measured slope of the superstable geometry is $\gamma\simeq 0.5$, consistent with a quadratic maximum ($z=2$) of the effective map (Fig.~\ref{fig: Poisson_nonlinearity}).

\begin{figure}[h]
\centering
\includegraphics[width=0.48\textwidth]{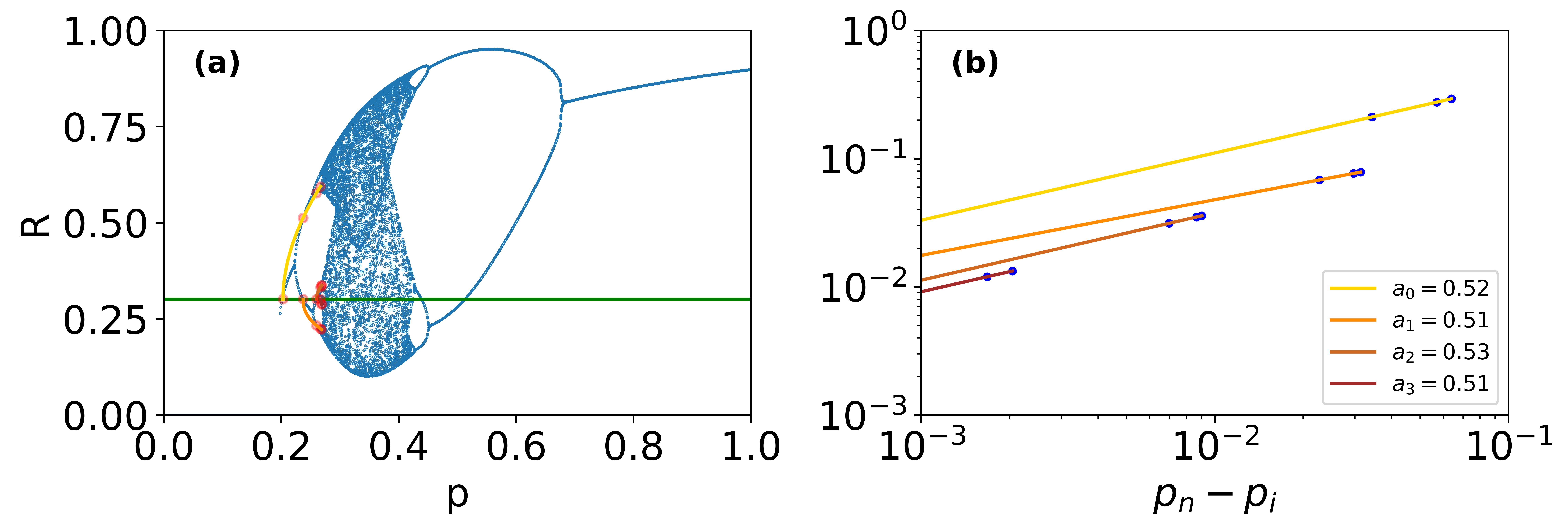}
\caption{Superstable geometry in triadic percolation (Poisson regulators). (a) Alternating-branch selection produces a self-similar pattern anchored at the critical height $R_m$. (b) Power-law fits of $|\mathcal{R}_n(p_n)-R_m|$ versus parameter distance yield $\gamma\simeq 0.5$, consistent with a quadratic local maximum. At each $2^n$–superstable parameter $p_n$ the tracked point $\mathcal{R}_n(p)$ coincides with the unique preimage that maps to $R_m$ in one iterate; away from $p_n$ it is its smooth continuation. $a_i$ is the slope from a least-squares fit of the data points $(\log \Delta p_{ni},\log d_{ni})$ over the chosen fit range (equivalently, the corresponding $i$-range) for the tracked branch/sequence.}
\label{fig: Poisson_nonlinearity}
\end{figure}

\subsection{Hill-type (effective) kernels: local quadratic dominance}
\label{sec:hill}
For intuition, we examine Hill-type effective kernels that yield closed forms for $p_L(x)$ without committing to a discrete degree PGF. For example,
\begin{equation}
p_L(x)=p\,(1-x)^{1/2}\bigl[1-(1-x)^{3/2}\bigr],
\end{equation}
which can be viewed as a convenient effective kernel with a similar shape to that obtained for regular structural networks, but is used here purely phenomenologically.
Over broad parameter ranges, log–log fits to the superstable sequence may show intermediate effective exponents ($0.15<\gamma<0.5$). However, as the accumulation region approaches, the local slope converges to $\gamma\to 0.5$, revealing a locally quadratic peak that dominates asymptotically (Fig.~\ref{fig: randomRegular})---as expected. This behavior also appears for the map
\begin{equation}
p_L(x)=p (1 - x)^{1/2}\bigl(1-(1 - x)^{7/2}\bigr),
\end{equation}
which is formally of order 4. However, because $p_L''(x)\neq 0$, the local peak remains quadratic and ultimately dominates the asymptotic behavior. Formally, the local nonflat order in the maximizer $x_m$ is the smallest $k\ge2$ with $F^{(k)}(x_m)\neq 0$, and here $F''(x_m)\neq0$, and hence $z=2$ asymptotically. Thus, a globally higher-order structure does not alter the local quadratic control of the onset. Empirically, extensive chaotic windows were more readily produced by noninteger exponents in these effective kernels than by simple integer-PGF choices; we do not claim impossibility for the latter, only that our exploration found the former to be a convenient and robust route.\\

\begin{figure}[t]
\centering
\includegraphics[width=0.48\textwidth]{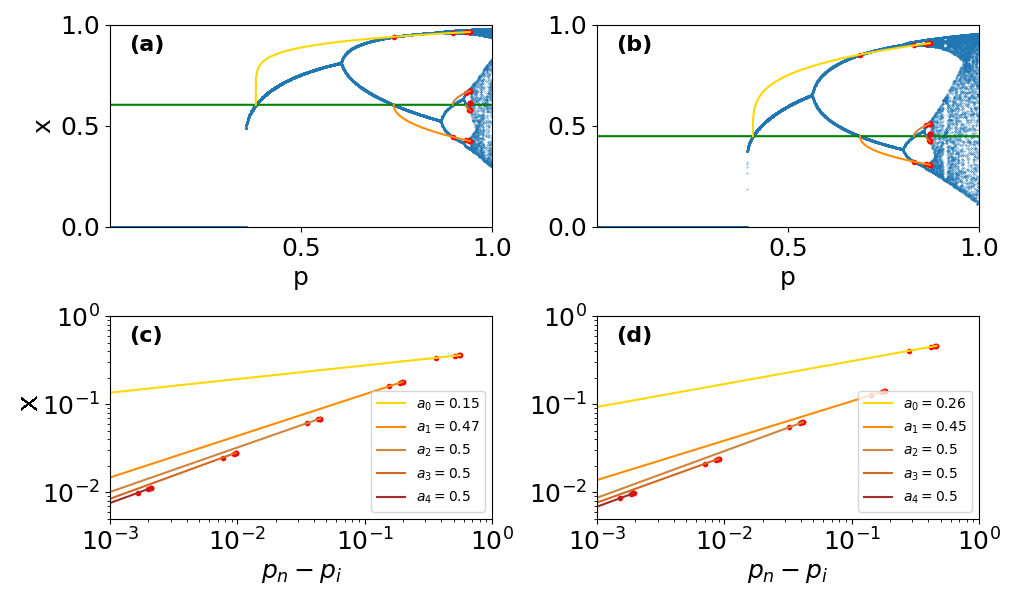}
\caption{Hill-type effective kernels. (a,b) Orbit diagrams and alternating-branch tracking for the kernels $p_L(x)=p\,(1-x)^{1/2}\bigl(1-(1-x)^{3/2}\bigr)$ and $p_L(x)=p\,(1-x)^{1/2}\bigl(1-(1-x)^{7/2}\bigr)$. (c,d) Broad-range fits can show intermediate slopes, but near the accumulation region the local exponent converges to $\gamma\to 0.5$, indicating local quadratic control. $a_i$ is the slope from a least-squares fit of the data points $(\log \Delta p_{ni},\log d_{ni})$ over the chosen fit range (equivalently, the corresponding $i$-range) for the tracked branch/sequence.}
\label{fig: randomRegular}
\end{figure}

\subsection*{Designing $z>2$ via regulator statistics (realization in triadic percolation)}
A natural question is whether random regular networks can produce triadic–percolation dynamics with a non-flat order $z\neq 2$ at the dynamical maximum. Using the activation kernel $F(x)$ (and $\phi(x):=\log F(x)$), the local order $z$ at an interior maximizer $x_m\in(0,1)$ is determined by
\begin{equation}
\phi^{(1)}(x_m)=\cdots=\phi^{(z-1)}(x_m)=0,\qquad \phi^{(z)}(x_m)<0,
\end{equation}
equivalently $F^{(k)}(x_m)=0$ for $1\le k\le z-1$ and $F^{(z)}(x_m)<0$. For generic choices of regulator statistics this yields $z=2$. Realizing $z>2$ therefore requires tuning the regulator statistics and/or the activation rule so that lower derivatives cancel at $x_m$ while the peak remains non-flat.

One illustrative construction is to shape the link-activation rule so that near the operating range the kernel behaves like a $d$th-order cap. For example, taking
\begin{equation}
p_L(x)=p\,\bigl(1-G_0^-(x)\bigr)
\end{equation}
and choosing a regular negative-regulator out-degree (so $G_0^-(x)=x^{d}$) gives
\begin{equation}
p_L(x)=p\,\bigl(1-x^{d}\bigr).
\end{equation}
This sets the local nonlinearity scale to $d$ (hence $z=d$ in the absence of lower-order contributions). To realize an interior maximizer $x_m\in(0,1)$ with the desired order $z>2$, one can combine inhibitory with or without excitatory regulators so that the product structure in $F(x)$ both places the maximum in the interior and enforces the cancellations $F^{(2)}(x_m)=\cdots=F^{(z-1)}(x_m)=0$ with $F^{(z)}(x_m)<0$. Alternative parameterizations,
\begin{equation}
p_L(x)=p\Bigl(1-\bigl((1-\theta)+\theta\,x^{d}\bigr)\Bigr)
\quad\text{or}\quad
p_L(x)=p_1-p_2\,x^{d},
\end{equation}
offer additional degrees of freedom to place $x_m$ in $(0,1)$ and cancel lower derivatives, thereby, at least in principle, enabling the engineering of cases with $z>2$ within the triadic–percolation framework.

Figures~\ref{fig: nonlinearity_2and4_triadicPercolation}(a,b) illustrate these ideas: panel (a) shows examples with regular negative regulation ($d=2$ and $d=4$) on a Poisson structural network; panel (b) shows a mixture model controlled by $\theta$. While simplified nonlinearities may hinder the appearance of long period-doubling cascades or chaos, the orbit geometry near the transition distinguishes the universality classes, with $z=d$ controlling the local order.

\subsection*{Unimodality and local quadratic form of the activation kernel}
Using the activation kernel $F(x)$ introduced above (with $f$ and $g$ as defined there), we collect a sufficient condition ensuring a single interior maximizer and a generically quadratic peak.

\emph{Unimodality (sufficient condition).} If $f$ is nondecreasing and $1-g$ is nonincreasing on $[0,1]$, and if $\log f$ and $\log\!\big(1-g\big)$ are concave on $[0,1]$ (e.g., for Poisson regulators or Hill-type $x^\alpha$, $1-x^\beta$ kernels), then
\begin{equation}
\frac{d}{dx}\log F(x)\;=\;\frac{f'(x)}{f(x)}\;-\;\frac{g'(x)}{1-g(x)}
\end{equation}
is nonincreasing. Hence it has at most one zero, and $F$ is unimodal with a unique maximizer $x_m\in(0,1)$. Under the same hypotheses (and smoothness), one generically has $F''(x_m)\neq 0$, i.e., a locally quadratic peak. Consequently, the local non-flat order at the maximum is $z=2$ in generic cases; deviations ($z>2$) require tuned cancellations of lower derivatives as discussed in the previous subsection.

\begin{figure}[t]
\centering
\includegraphics[width=0.48\textwidth]{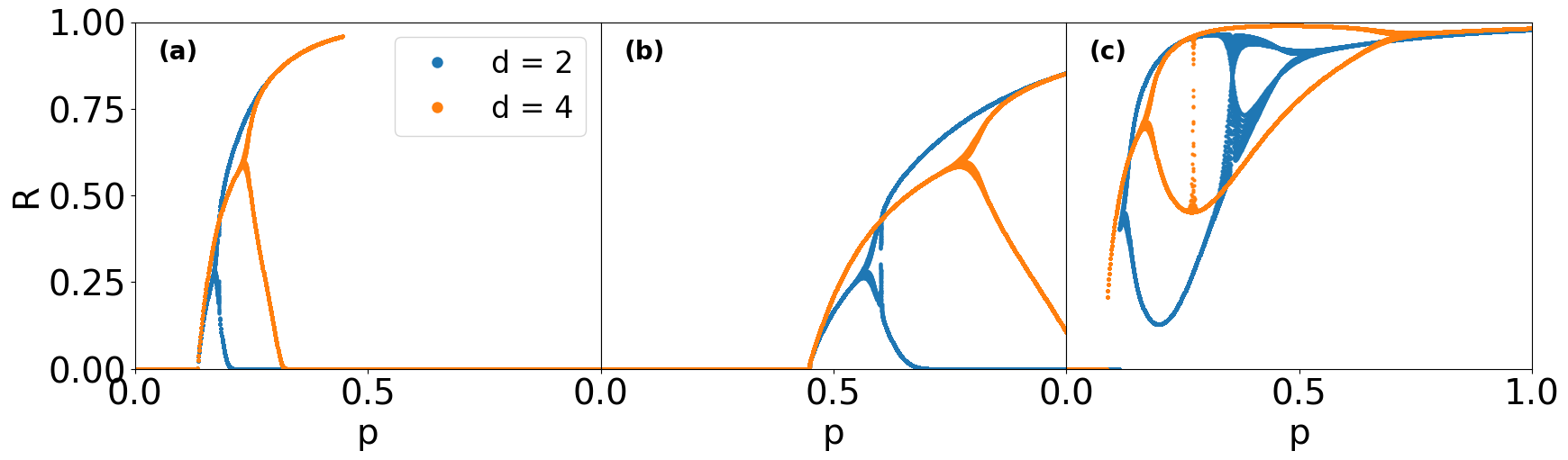}
\caption{Examples of triadic-percolation dynamics under simplified regulatory actions controlling the system’s nonlinearity (“cap”) set to quadratic or quartic. In all three cases, the structural network is generated from a Poisson distribution, with different $c$. 
(a) $c=7.5$: dynamics are regulated solely by negative regulators, with $p_L^{(t)} = p \bigl(1 - G_0^-(R^{(t)})\bigr)$. When $G_0^-$ is taken as a regular network of degree $d$, the system’s nonlinearity---and thus its local universality class in the sense of the $z$-logistic unimodal maps---is controlled by $d$.  Since there is no period doubling or chaotic regime, the geometry of superstable points does not apply; however, the orbit diagram near the transition suggests a universality different from the classical ($d=2$) when $d=4$. 
(b) Same as in (a), except the regulator is a mixture of two regular networks: $p_L^{(t)} = p \bigl(1 - \bigl((1-\theta) + \theta\,G_0^-(R^{(t)})\bigr)\bigr)$ with $\theta=0.3$. 
(c) System with $c=22$ and $p_L^{(t)} = p_1 - p_2 \bigl(R^{(t)}\bigr)^{d}$ for $p_1=0.58$ and $p_2=0.42$.}
\label{fig: nonlinearity_2and4_triadicPercolation}
\end{figure}

\section{Conclusion}
Triadic percolation turns bond percolation into a dynamical problem, whose long–time behavior is governed by an effective one–dimensional map. We find that the geometry of superstable windows provides a map–agnostic probe of the local nonlinearity at the dynamical maximum: the distance of the next–to–maximum point on the cycle obeys $|r_n-R_m|\propto(\Delta p)^{\gamma}$ with $\gamma=1/z$, where $z$ is the nonflat order of the peak. Numerically, this law holds across canonical unimodal families and across triadic percolation with heterogeneous structural and regulatory ensembles. Lyapunov spectra corroborate effective one–dimensionality.

This diagnostic acts directly on data---without reconstructing the governing map---and benefits from forward/inverse cascades that widen usable fitting windows. Its scope is local: under unimodality, smoothness, and a nonflat interior maximum the slope is $1/z$, as in standard unimodal-map universality, whereas multiple critical points, nonunimodal kernels, or nonlinear parameterizations can alter the exponent; in such cases explicit branch selection and derivative checks can identify the correct local class. Looking ahead, extracting $z$ from experimental orbit diagrams in biological, social, and technological systems with regulatory triads, extending the framework to asymmetric or multifold critical points and to correlated regulator statistics, and combining geometry–based diagnostics with renormalization–style scalings may help delineate universality boundaries in higher–order interaction models.

Beyond standard unimodal families, the same diagnostic can be used whenever one has orbit-diagram-like state--parameter data from a controlled parameter sweep,
even if the effective map is not known in closed form. For example, recent studies of percolation-driven chaos in networked/higher-order systems provide such
parameter-sweep data (see the added references suggested by the referee). In such settings one can locate superstable windows via $\lambda(p)$ and extract $\gamma$
from the measured approach of the distinguished branch toward $R_m$.

\paragraph*{Acknowledgments.}
This work was funded by the Deutsche Forschungsgemeinschaft (DFG, German Research Foundation)---Project No.~557852701 (A.A.S.).

\bibliography{Refs}

\end{document}